%% file: ActiveNematics_Rev.tex
\documentclass[aps,pre,twocolumn,nofootinbib,floatfix]{revtex4-1}

\usepackage{amsmath,amssymb,mathrsfs,bm}
\usepackage{color,graphicx,subfigure}
\usepackage[english]{babel} 
\newcounter{pp}[section]

\newcommand{\bB}{\mathbf{B}}
\newcommand{\bb}{\mathbf{b}}
\newcommand{\bBe}{\mathbf{B}_{\mathrm{e}}}

\newcommand{\bbD}{\mathbb{D}}

\newcommand{\bD}{\mathbf{D}}

\newcommand{\be}{\mathbf{e}}
\newcommand{\bF}{\mathbf{F}}
\newcommand{\FF}{\mathcal{F}}
\newcommand{\bFe}{\mathbf{F}_{\mathrm{e}}}

\newcommand{\bg}{\mathbf{g}}
\newcommand{\bG}{\mathbf{G}}
\newcommand{\bh}{\mathbf{h}}
\newcommand{\bH}{\mathbf{H}}
\newcommand{\bI}{\mathbf{I}}
\newcommand{\bbI}{\mathbb{I}}

\newcommand{\bPsi}{\bm{\Psi}}

\newcommand{\bmm}{\mathbf{m}}
\newcommand{\bn}{\mathbf{n}}
\newcommand{\bno}{\mathring{\mathbf{n}}}

\newcommand{\bQ}{\mathbf{Q}}

\newcommand{\bt}{\mathbf{t}}

\newcommand{\bT}{\mathbf{T}}
\newcommand{\bTa}{\mathbf{T}_{\text{a}}}
\newcommand{\bTz}{\mathbf{T}_{\zeta}}
\newcommand{\bW}{\mathbf{W}}
\newcommand{\bw}{\mathbf{w}}

\newcommand{\bv}{\mathbf{v}}

\newcommand{\bX}{\mathbf{X}}

\renewcommand{\rho}{\varrho}
\newcommand{\eps}{\varepsilon}
\newcommand{\si}{\sigma}
\newcommand{\sio}{\sigma_{0}}

\newcommand{\uctd}{\triangledown}

\newcommand{\taur}{\tau_{\text{rel}}}
\newcommand{\taudef}{\tau_{\text{def}}}

\newcommand{\Ldot}[1]{\overset{\bm{.}}{#1}}

\newcommand{\Pt}{\mathcal{P}_t}
\newcommand{\bnu}{\bm{\nu}}
\newcommand{\Wext}{W^{(\text{ext})}}

\DeclareMathOperator{\tp}{\otimes}

\DeclareMathOperator{\tr}{tr}

\DeclareMathOperator{\divr}{div}

\newcommand{\D}[2]{\frac{\partial #1}{\partial #2}}

\newcommand{\comm}[1]{{\color{black}#1}}

\newtheorem{lemma}{Lemma}

\begin{document}

\title{Active nematic gels as active relaxing solids}
\author{Stefano S.\ Turzi}

\affiliation{\mbox{Dipartimento di Matematica, Politecnico di Milano, Piazza Leonardo da Vinci 32, 20133 Milano, Italy}}

\date{\today}

\begin{abstract} I put forward a continuum theory for active nematic gels, 
defined as fluids or suspensions of orientable rodlike objects endowed with 
active dynamics, that is based on symmetry arguments and compatibility with 
thermodynamics. The starting point is our recent theory that models (passive) 
nematic liquid crystals as relaxing nematic elastomers. The interplay between 
viscoelastic response and active dynamics of the microscopic constituents is 
naturally taken into account. By contrast with standard theories, activity is 
not introduced as an additional term of the stress tensor, but it is added as 
an \emph{external remodeling force} that competes with the passive relaxation 
dynamics and drags the system out of equilibrium. In a simple one-dimensional 
channel geometry, we show that the interaction between non-uniform nematic 
order and activity results in either a spontaneous flow of particles 
or a self-organization into sub-channels flowing in opposite directions.
\end{abstract}

\maketitle

\section{Introduction}
\label{sec:intro}

The mathematical modeling of biophysical and active materials pose new 
theoretical challenges for their unique features. Examples of active matter 
comprise bacterial swarms \cite{09lubensky}, the cellular cytoskeleton 
\cite{07julicher,15Prost} and in vitro cell extracts. Non-biological examples 
include vibrated granular material \cite{13Marchetti}. These system have 
attracted much interest in recent years both from a purely theoretical 
perspective and for their potential applications. They are characterized by a 
strong deviation from thermal equilibrium due to the environmental energy supply 
and the active dynamics of the system's microscopic subunits. The simplest, yet 
successful, theoretical description of active matter is based on continuum 
models for single-component suspensions of rodlike objects. These models have 
been originally developed to describe (passive) nematic liquid crystals, e.g., 
Ericksen-Leslie theory. The key features of an active system, namely, its 
viscoelastic response \cite{15Prost,15Cates,16Cates} and the active dynamics due 
to energy consumption of the material sub-units, are usually added in an 
\emph{ad hoc} manner to the passive physical model. Viscoelasticity is taken 
into account by postulating a Maxwell relaxation time and activity is usually 
introduced by assuming an \emph{active stress}, proportional to the nematic 
ordering tensor $\bQ$. In its simplest form, the active stress is postulated to 
be of the form $\bTz=-\zeta \bQ$, where $\zeta$ is a modeling coefficient that 
measures the strength of activity. When $\zeta>0$ the material is extensile, 
while if $\zeta<0$ it has a natural tendency to contract. This contribution to 
the stress tensor has been deduced on microscopic grounds by Simha and Ramaswamy 
\cite{02Rama}, and it has been widely used since then (see, for instance, Refs. 
\cite{07julicher,13Marchetti,15Yeomans,15Prost,16Cates}). The equations that 
emerge are those proposed by Simha and Ramaswamy \cite{02Rama} for 
self-propelling organisms, but similar models have been developed in the 
context of the cytoskeleton of living cells, a network of polar actin filaments, 
made active by molecular motors that consume ATP \cite{07julicher,15Prost}. A 
multi-component theory based on irreversible thermodynamics is derived in 
\cite{05julicher,07julicher,16Forest}.  The theory seems to be able to make a 
number of successful predictions, e.g., the onset of spontaneous flow 
\cite{05voituriez,09edwards}, motility and spontaneous division of active 
nematic droplets \cite{11Amar,12Cates,14GiomiDS}.

However, I believe that the assumption of an \emph{active stress} does not 
correctly capture the true essential nature of an active behavior. The responses 
of tissues to elastic forces are quite different from the passive mechanical 
properties of composite materials. For example, on short time scales the passive 
elastic response of the matrix and the cellular cytoskeleton dominate the 
mechanical response of the tissue; on longer time scales many cell types (such 
as muscle cells, fibroblasts, endothelial cells) can reorganize to reduce their 
internal stress and thus reach a relaxed or natural state. Cross-links between 
polymer filaments define a natural distance, in other words they define a 
natural metric in the material, that I shall call \emph{shape tensor}. A 
strained viscoelastic material has a natural tendency to recover this natural 
state and can reconfigure its internal structure to perform this relaxation in 
an efficient manner (e.g. viscous relaxation).

By contrast, activity is an \emph{external remodeling force} that competes with 
the passive remodeling and may drive the microscopic reorganization away from 
the natural metric. The remodeling forces model the interaction with the 
chemical fuel, and tension is generated as a consequence of the crawling motion 
of aligned filaments. Hence, I believe that activity is best described by a term 
in the evolution equation of the internal structure of the material, rather than 
directly in the stress tensor. Indeed, chemical fuel is consumed, and hence the 
power exerted by the active term should be nonzero, even in the absence of any 
macroscopic flow. This argument contrasts with the introduction of the
active term $\bTz$ in the Cauchy stress tensor: if the macroscopic flow, 
$\bv$, is zero, the stress power, calculated as $\bTz \cdot \nabla \bv$, must 
also vanish.

It is worth noticing that, since the actual macroscopic response of the 
material is related to the ``distance'' between its present state and its 
relaxed state, activity has an indirect contribution to the stress tensor as the 
effective material response depends on it. I will make these statements 
mathematically more precise in the next Sections. 

A similar situation occurs in the continuum theory for growth and remodeling of 
solid-like biological tissues, e.g., muscles 
\cite{05adc,07Ambrosi,08Holzapfel,09Preziosi,10Ambrosi,12Ambrosi,17Teresi}. Some 
theories use a remodeling force and an active strain (instead of an active 
stress) to model the ability of the muscle to actively modify its natural 
metric. The remodeling is then performed, via a remodeling force, at the 
expenses of a chemical fuel. Chemical energy must be supplied also in the 
absence of any macroscopic motion, for example during an isometric exercise, 
as it is needed for the remodeling of the muscle internal structure. Hence, to 
calculate the stress power, the active term has to be paired with a kinematic 
quantity related to the material remodeling rather than with the macroscopic 
velocity gradient $\nabla \bv$.

An active stress of the form $\bTz = -\zeta \bQ$ has also been criticized by 
Brand, Pleiner and Sven{\v{s}}ek \cite{14Brand} on the ground of compatibility 
with linear irreversible thermodynamics. Their argument is related to the fact 
that reversible and dissipative forces behave differently under time-reversal: 
the dissipative forces have the same signature as their conjugate fluxes under 
time-reversal, while reversible forces have the opposite signature. 
Time-reversal implies a change of sign of the velocity field $\bv$. In our case, 
the active stress is coupled with the gradient of velocity $\nabla \bv$ and its 
contribution to the entropy production is $-\zeta \bQ \cdot \nabla \bv$. If 
$\bTz$ has to be irreversible, then the activity coefficient $\zeta$ must also 
change sign under time reversal, which is a rather unusual behavior for a 
scalar.

In the present paper we propose a simple and systematic derivation of the 
continuum equations of viscoelastic active nematic gels that is based on 
symmetry arguments, relaxation dynamics and compatibility with linear 
irreversible thermodynamics. For simplicity, we discuss only a one-component 
active gel considering therefore that the complex composition of active 
materials such as the cytoskeleton can be described by an effective single 
component fluid. More complicated theories that employ a multi-component 
description of the active fluid and use linear irreversible thermodynamics are 
to be found in Refs.\cite{05julicher,16Forest}. 

The theory is developed in Sec.\ref{sec:theory}. Sec.\ref{sec:weak-flow} 
describes the hydrodynamic approximation, and Sec.\ref{sec:SpontaneousFlow} 
deals with some simple applications of the theory.

\section{Shape tensor and relaxation dynamics}
\label{sec:theory}
Since continuum theories are essentially based on symmetry arguments and general 
physical principles, they are generally applicable to a whole range of physical 
systems. On the downside, they do not involve significant microscopic 
considerations and the material parameters are purely phenomenological. In the 
present Section we put forward a continuum theory for nematic active gels. The 
viscoelastic passive response is derived by relaxing the elastic response of a 
nematic liquid crystal elastomer. The shape tensor in this theory plays the role 
of a metric tensor and describes, on a macroscopic ground, the information about 
the equilibrium distances among the centers of mass of the microscopic 
constituents.

\subsection{Internal and external degrees of freedom}
The first key idea is to study, separately, the degrees of freedom associated 
with the elastic deformations and the ``microscopic'' degrees of freedom 
related to the material relaxation and reorganization. The first process is 
reversible and conserves the energy, while the second process involves material 
reorganization and is irreversible. I assume that activity directly interferes 
with the \emph{internal degrees} of freedom and competes with the natural 
tendency of the material to reach the equilibrium state. To this end, we 
introduce the Kr\"{o}ner-Lee-Rodriguez multiplicative decomposition for the 
deformation gradient $\bF=\bFe \bG$ \cite{59Kroner,69Lee,94Rodriguez}. For later 
convenience we also define the inverse relaxing strain $\bH = 
(\bG^{T}\bG)^{-1}$, so that the \emph{effective} left-Cauchy-Green deformation 
tensor can be written as
\begin{equation}
\bBe = \bFe\bFe^{T} =\bF\bG^{-1} \bG^{-T}\bF^{T} = \bF \bH \bF^{T}.
\end{equation}
The same decomposition has been recently applied to explain the hints of 
viscoelasticity that remain at the hydrodynamic level when a sound wave 
propagates inside a nematic crystals \cite{14bdt,15tur,16bdt,16Turzi}.

The tensor $\bG$ models the \emph{microscopic remodeling} of the material. In 
other words, it describes the inelastic dissipative (irreversible) processes 
within the material. By contrast, $\bFe$ is related to the \emph{elastic 
(reversible) response}.

\subsection{Shape tensor and free energy}
Contrary to most hydrodynamic theories of active nematic gels I do not use the 
ordering tensor $\bQ$ to take into account uniaxial nematic symmetry of the 
microscopic sub-units. I rather introduce a uniaxial, unit determinant, 
\emph{shape tensor}, common to the theory of nematic elastomers 
\begin{align}
\bPsi(\rho,\bn) & = a(\rho)^2 (\bn \tp \bn) + a(\rho)^{-1}\big(\bI - \bn \tp \bn \big),
\end{align}
where $a(\rho)$ is a (density dependent) shape parameter and the preferred 
direction $\bn$ lives in the actual configuration of the body since it is not 
materially linked to body deformations. \comm{The shape tensor is spherical, prolate or oblate respectively for $a(\rho) 
= 1$, $a(\rho) > 1$ or $a(\rho) < 1$. The material parameter $a(\rho)$ gives the amount of spontaneous elongation along 
$\bn$ in a uniaxially ordered phase. It is a combined measure of the degree of order and of the strength of the 
nematic-elastic coupling. The tensor $\bPsi$ represents a volume-preserving uniaxial stretch along the current direction 
of the director $\bn$. In particular, the unit determinant assumption implies that growth is not taken into account in 
the model.}

It must be noted that the same tensor is used to describe the coupling between strain and orientation in nematic 
elastomers, \comm{where $\bPsi$ is usually interpreted as a effective step-length tensor that reflects 
the current nematic ordering in the polymer network \cite{Warner}. In our model, $\bPsi$ represents the 
spontaneous metric tensor that is dictated by the coarse-grained anisotropy of the subunits. The equilibrium 
configuration of the sub-units is usually anisotropic in the direction of $\bn$ and the measure of this anisotropy is 
yielded by the value of $a(\rho)$.}

For fast relaxation times, only the local form of the elastic energy is 
important, so we don't need to specify its exact expression globally. However, 
in this context it is natural to assume the standard energy of polymer physics, 
i.e., neo-Hookean elasticity. In particular, given the uniaxial symmetry of the 
constituents, I posit that the elastic response is governed by the nematic 
elastomer free energy \cite{Warner}, written in terms of $\bFe$. Hence, I posit 
the following free energy density per unit mass

\begin{equation}
\begin{split}
\si(\rho,\bBe,& \bn,\nabla\bn) = \sio(\rho) +  \tfrac{1}{2}\mu \Big(\tr \big(\bPsi^{-1}\bBe - \bI \big) \\
& - \log\det\big(\bPsi^{-1}\bBe \big)\Big)
+ \sigma_{\text{Fr}}(\rho,\bn, \nabla\bn),
\end{split}
\label{eq:DensitaEnergiaPsi}
\end{equation}
where $\rho$ is the density, $\rho \mu$ is the shear modulus, and $\bI$ is the 
identity tensor. I have also introduced the classical Oseen-Frank potential 
$\sigma_{\text{Fr}}(\rho,\bn, \nabla\bn)$ \cite{95dgpr} that favors the 
alignment of the director field $\bn$. The isotropic term, $\sio(\rho)$, takes 
into account compressibility. It does not depend on $\bFe$ and is thus not 
affected by stress relaxation. This is related to the fact that stresses do not 
vanish in a purely isochoric deformation. By contrast, viscoelastic material 
relax the \emph{shear-stress} after a sufficiently long time.

\subsection{Dissipation}
Let $\Pt$ be an arbitrary region that convects with the body. We restrict 
attention to a purely mechanical theory based on the requirement that the 
temporal increase in kinetic and free energy of $\Pt$ be less than or equal to 
the power expended on $\Pt$ by the external forces. The difference being the 
power dissipated in irreversible processes. Specifically, for any isothermal 
process, for any portion $\Pt$ of the body at all times, we require

\begin{align}
\mathcal{D} := \Wext - \dot{K} - \dot{\FF} \geq 0 , 
\end{align}
where $\Wext$ is the power expended by the external forces, $\dot{K}$ is the 
rate of change of the kinetic energy, $\dot{\FF}$ is the rate of change of the 
free energy, and the dissipation $\mathcal{D}$ is a positive quantity that 
represents the energy loss due to irreversible process (entropy production). 
Here, an overdot indicates the material time derivative. More precisely, I 
define 

\begin{align}
\Wext & := \int_{\Pt} \bb\cdot \bv \,\,dv + \int_{\partial\Pt} \bt_{(\bnu)}\cdot \bv \,\,da \notag \\
& + \int_{\Pt} \bg \cdot \dot{\bn} \,\,dv + \int_{\partial\Pt} \bmm_{(\bnu)} \cdot \dot{\bn} \,\,da \notag \\
& + \int_{\Pt} \bTa\cdot \bBe^{\uctd} \,\,dv, \label{eq:Wext} \\
K + \FF & := \int_{\Pt} \left(\frac{1}{2}\rho \bv^2 + \rho \si(\rho,\bBe,\bn,\nabla\bn) \right) \,dv, \\
\mathcal{D} & = \int_{\Pt} \xi \,\,dv, \qquad \xi \geq 0,
\end{align}
where $\bv$ is the velocity field, and 
\begin{equation}
\begin{split}
\bBe^{\uctd}& := (\bBe)\Ldot{\phantom{I}} - (\nabla\bv) \,\bBe - \bBe \,(\nabla\bv)^T \\
&= \bF\Ldot{\bH}\bF^{T},
\end{split}
\label{eq:Be_codeform}
\end{equation}
is the codeformational derivative\footnote{\comm{Also known as upper-convected time derivative, upper-convected rate or 
contravariant rate.}} \cite{Joseph,Gurtin}, a frame-indifferent time-derivative of $\bBe$ 
relative to a convected coordinate system that moves and deforms with the 
flowing body. The unit vector $\bnu$ is the external unit normal to the boundary 
$\partial\Pt$; $\bb$ is the external body force,  $\bt_{(\bnu)}$ is the external 
traction on the bounding surface $\partial\Pt$. The vector fields $\bg$ and 
$\bmm_{(\bnu)}$ are the external generalized forces conjugate to the 
microstructure: $\bn \times \bg$ is usually interpreted as ``external body 
moment'' and $\bn \times \bmm_{(\bnu)}$ is interpreted as ``surface moment per 
unit area'' (the couple stress vector). This interpretation comes from the 
identity $\dot{\bn} = \bw \times \bn$, where $\bw$ is the (local) angular 
velocity of the director, so that, for instance, the external power density is 
written as $\bg \cdot \dot{\bn} = \bw \cdot (\bn \times \bg)$.

The last term in Eq.\eqref{eq:Wext} is particularly interesting and new: $\bTa$ 
is a second-rank tensor that represents an \emph{external remodeling force} 
\cite{02DiCarlo}, i.e., an external \comm{generalized force} that competes with the natural 
microscopic reorganization of the body. It must be noted that $\bTa$ \comm{has the same dimensions of a Cauchy 
stress tensor} and is conjugate to the remodeling velocity field $\bBe^{\uctd}$. By contrast, the classical 
active stress $\bTz$ is paired with the macroscopic velocity gradient 
$\nabla\bv$. The possibly unfamiliar time-derivative $\bBe^{\uctd}$ has the 
right properties to represent the kinematics of reorganization: (1) it is 
frame-invariant, (2) it vanishes whenever the deformation is purely elastic and 
there is no evolution of the natural configuration\footnote{\comm{It is clear from Eq.\eqref{eq:Be_codeform} that 
$\bBe^{\uctd} = 0$ if and only if $\Ldot{\bH}=0$. The tensor $\bH = (\bG^{T}\bG)^{-1}$ is related to material remodeling 
and no remodeling occurs when the deformation is purely elastic.}}, and (3), as shown in the 
Appendix, it comes out naturally when studying the passive remodeling (see 
Eq.\eqref{eq:dotF} of the Appendix, where the 
material time-derivative of $\FF$ is explicitly calculated). The same derivative also appears in the 
three-dimensional models for Maxwell viscoelastic fluids \cite{Joseph}. Finally, the remodeling power, $\bTa\cdot 
\bBe^{\uctd}$, depends only on the 
point value of the internal velocity field $\bBe^{\uctd}$, and not on any of its 
spatial gradients. Hence, as far as remodeling dynamics is concerned, the 
present is therefore a theory of grade zero.

After some algebra, reported in the Appendix for ease of reading, the dissipation is recast in the following form
\begin{align}
\mathcal{D} & = \int_{\Pt} \left(\bb - \rho \dot{\bv} + \divr\bT \right) \cdot \bv \,\,dv \notag \\
& + \int_{\partial\Pt} \left(\bt_{(\bnu)} - \bT\bnu \right) \cdot \bv \,\,da 
+ \int_{\Pt} \left(\bg - \bh \right) \cdot \dot{\bn} \,\,dv \notag \\
& + \int_{\partial\Pt} \left(\bmm_{(\bnu)}  - \left(\rho\D{\si}{\nabla\bn}\right) \bnu \right) \cdot \dot{\bn} \,\,da \notag \\
& + \int_{\Pt}\left(\bTa-\rho\D{\si}{\bBe}\right)\cdot \bBe^{\uctd}\,\, dv,
\label{eq:dissipazioneIntegrale}
\end{align}
where the Cauchy stress tensor and the molecular field are found to be (again for more details, see the Appendix)
\begin{align}
\bT & = -\rho^2\D{\si}{\rho}\bI + 2\rho\D{\si}{\bBe}\bBe - \rho(\nabla \bn)^T\D{\si}{\nabla\bn}, 
\label{eq:Tdef} \\
\bh & := \rho\D{\si}{\bn} - \divr\left(\rho\D{\si}{\nabla\bn} \right). \label{eq:hdef} 
\end{align}

\subsection{Governing equations}
According to the model, the material response is elastic with respect to the 
natural configuration. In other words, energy dissipation is uniquely associated 
to the evolution of the natural or stress-free configuration of the body, i.e., 
energy is dissipated only when microscopic reorganization occurs. As a 
consequence, only the term containing $\bBe^{\uctd}$ in 
Eq.\eqref{eq:dissipazioneIntegrale} yields a positive contribution to the 
dissipation, while the first four integrals must vanish. Given the arbitrariness 
of $\Pt$ and of the test fields, it is natural to use a generalized Rayleigh 
principle \cite{01sovi,12Virga_book} and impose the vanishing of the 
corresponding generalized forces. In our case this yields the usual balance of 
momentum equation

\begin{equation}
\rho \dot{\bv} = \bb + \divr\bT 
\end{equation}
with boundary condition $\bt_{(\bnu)} = \bT\bnu$. In particular, when the energy 
density \eqref{eq:DensitaEnergiaPsi} is substituted into Eq. \eqref{eq:Tdef}, 
the Cauchy stress tensor reads
\begin{align}
\bT & = -p \,\bI + \rho \mu \big(\bPsi^{-1}\bBe - \bI \big) - \rho(\nabla \bn)^T\D{\si}{\nabla\bn},
\label{eq:Tstress}
\end{align}
where the pressure-like function $p$ is
\begin{align}
p & = \rho^2 \Big[\D{\sio}{\rho} -\mu \frac{3 a'(\rho)}{2a(\rho)} \left(\bn\tp\bn-\frac{1}{3}\bI\right) \cdot \big(\bPsi^{-1}\bBe\big) \Big].
\label{eq:pressure}
\end{align}
It is worth noticing that $\bT$ does not explicitly contain any ``active 
component''. Activity is taken into account implicitly via the evolution 
equation for the natural configuration that describes how $\bBe$ evolves in 
time. I shall describe this equation below. 

To obtain an equation for the microstructure $\bn$, we impose the vanishing of 
the third and fourth integrals in Eq. \eqref{eq:dissipazioneIntegrale}. Since 
$\bn \cdot \Ldot{\bn} = 0$, it is sufficient to posit
\begin{equation}
\bn \times \big(\bg - \bh \big) = 0,
\label{eq:evoluzione_direttore} 
\end{equation}
with boundary condition
\begin{equation}
\bn \times \bmm_{(\bnu)} = \bn \times \left(\rho\D{\si}{\nabla\bn}\right) \bnu.
\end{equation}
These two last equations describe the evolution of the director field and are 
usually interpreted as balance of torques. Contrary to the first appearance they 
do not describe a dissipation-less motion, energy dissipation is again contained 
in their dependence on $\bBe$. It is possible to show that, under the 
approximation of fast relaxation times, Eq.\eqref{eq:evoluzione_direttore} 
reproduces exactly the usual director dynamics of liquid crystals, where the 
rotational Leslie coefficient $\alpha_2$ and $\alpha_3$ are identified in terms 
of our model parameters \cite{14bdt,16Turzi}.

Finally, a positive dissipation for any $\Pt$ at all times implies 
\begin{equation}
\left(\bTa-\rho\D{\si}{\bBe}\right)\cdot \bBe^{\uctd} \geq 0,
\label{eq:dissp_inequality}
\end{equation}
where in our specific case
\begin{align}
\D{\si}{\bBe} = \tfrac{1}{2}\mu \big(\bPsi^{-1} - \bBe^{-1}\big).
\label{eq:Ds/DBe}
\end{align}
It is customary, when dealing with irreversible processes near equilibrium, to 
interpret the dissipation (or rate of entropy production) as the product of 
``fluxes'' and ``forces'' and to assume a linear coupling between them. 
Furthermore, Onsager reciprocal relations impose additional symmetry constraints 
on this linear dependence. Hence, I assume that the evolution of the 
microscopic remodeling is governed by the following ``gradient-flow'' equation 
for $\bBe$
\begin{equation}
\bbD (\bBe^{\uctd}) + \rho\D{\si}{\bBe} = \bTa \, ,
\label{eq:EquazioneEvoluzionePsi}
\end{equation}
where $\bbD$ is a fourth-rank tensor with the major symmetries. It is positive 
definite so that the dissipation inequality \eqref{eq:dissp_inequality} is 
automatically satisfied. Moreover, I take it to be compatible with the 
underlying uniaxial symmetry along $\bn$ of the microscopic constituents. The 
elements of the dissipation tensor $\bbD$ contain the characteristic relaxation 
times, and it is possible to show that there are only four relaxation times 
allowed by symmetry for rod-like constituents. I refer the interested reader to 
Ref.\cite{16Turzi} for a more detailed discussion on these points.

\subsection{Interpretation of the active stress}
By contrast with standard theories, the present model does not explicitly 
include an active term in the Cauchy stress tensor. However, it is possible to 
recast our stress tensor, as given in Eq.\eqref{eq:Tdef}, in a different form so 
that an analogue of the ``active stress'' appears. To this end, I substitute 
Eq.\eqref{eq:EquazioneEvoluzionePsi} into Eq.\eqref{eq:Tdef} and obtain

\begin{align}
\bT = -\rho\D{\si}{\rho}\bI  + 2\left(\bTa-\bbD (\bBe^{\uctd}) \right)\bBe - \rho(\nabla \bn)^T\D{\si}{\nabla\bn}.
\label{eq:Tdef_active}
\end{align}
The active contribution now comes from the term $2\bTa\bBe$ and it is not 
exactly equivalent to $\bTz$ because it contains the effective strain tensor 
$\bBe$. However, the analogy becomes more concrete if we make the approximation 
of small effective deformations, studied in the next Section.

\section{Small effective deformations}
\label{sec:weak-flow}
There are essentially two time-scales in the problem. One characteristic time is 
dictated by the macroscopic deformation and is thus related to $\nabla \bv$ in 
the following way

\begin{equation}
\taudef=1/\|\nabla \bv\|. 
\end{equation}
The second characteristic time is connected to material remodeling and 
determines the time-rate at which $\bBe$ reaches its equilibrium value. I posit 
$\taur = 2\|\bbD\|/\rho\mu$, where $\rho\mu$ is a characteristic shear modulus. 

We want to study the asymptotic approximation of the theory in the limit $\taur \ll \taudef$, that is
\[\taur \|\nabla \bv\| = \eps \ll 1. \]
In this limit reorganization is much faster than deformation and the theory 
reduces to a purely hydrodynamic theory, i.e., viscoelasticity becomes a higher 
order phenomenon and, to first order, viscosity coefficients are obtained as the 
product of the shear modulus and the relaxation times. An analysis of this 
approximation for the passive case is presented in Ref.\cite{16bdt,16Turzi}

For simplicity, I further assume that the active term only introduces a small 
perturbation of the passive dynamics so that the effective strain tensor  $\bBe$ 
is only a slight perturbation of its (passive) equilibrium value
\begin{align}
\bBe = \bPsi + \bB_1,\qquad \text{ with } \quad \|\bB_1\| = O(\eps),
\label{eq:approx_lin}
\end{align}

To leading order, the substitution of \eqref{eq:approx_lin} into Eq. \eqref{eq:Tdef_active} yields the Cauchy stress-tensor
\begin{align}
\bT & = -\rho^2\D{\si}{\rho}\bI  - 2\bbD (\bPsi^{\uctd})\bPsi 
+ 2\bTa\bPsi \notag \\
& - \rho(\nabla \bn)^T\D{\si}{\nabla\bn}.
\label{eq:Tdef_active_weak}
\end{align}
The codeformational derivative of the shape tensor reads
\begin{align}
\bPsi^{\uctd} & = \frac{\rho a'(\rho)}{a(\rho)^2} (\tr\bD) \left(\bI - (1+2a(\rho)^3)(\bn\tp\bn) \right) \notag \\
& + \big(a(\rho)^2 - a(\rho)^{-1} \big)\big(\bno\tp\bn + \bn \tp \bno \notag \\
& - \bD\bn\tp\bn- \bn\tp\bD\bn \big) 
- 2a(\rho)^{-1}\,\bD ,
\label{eq:bLup}
\end{align}
where $\bno = \Ldot{\bn} - \bW\bn$, $\bW=(\nabla\bv - \nabla\bv\,^T)/2$ is the 
spin tensor and $\bD=(\nabla\bv + \nabla\bv\,^T)/2$ is the stretching tensor.

The first two terms in Eq.\eqref{eq:Tdef_active_weak} reproduce the passive 
dynamics of nematic liquid crystals. Indeed, they reduce exactly to the Cauchy 
stress tensor as given in the classical (compressible) Ericksen-Leslie theory 
\cite{95dgpr}
\begin{equation}
\begin{split}
\bT_{\text{EL}} & = -p\bI + \alpha_1 (\bn\cdot \bD\bn)(\bn\tp\bn) \\
& + \alpha_2 (\bno \tp  \bn) + \alpha_3 (\bn \tp  \bno) \\
& + \alpha_4 \bD + \alpha_5 (\bD\bn \tp \bn) + \alpha_6 (\bn \tp \bD\bn) \\ 
& + \alpha_7 \big((\tr \bD ) (\bn\tp\bn) + (\bn\cdot \bD\bn)\bI \big)\\
& + \alpha_8 (\tr \bD ) \bI,
\end{split}
\label{eq:TLeslie}
\end{equation}
after the viscosity coefficients $\alpha_1, \ldots, \alpha_8$ have been suitably 
identified. An explicit comparison is carried out in \cite{16Turzi}. 

The third term, $2\bTa\bPsi$, \comm{corresponds to an \emph{active stress} and as such can be compared with the 
active stress $\bTz \propto \bQ$, as given in the usual theories 
\cite{02Rama,12Giomi,13Marchetti,15Prost,15Yeomans,16Cates}. With the simple choice
\begin{align}
\bTa = -\tfrac{1}{2}\rho \mu \,\zeta \,\bI,
\end{align}
where $\zeta$ is a dimensionless modeling parameter that controls the activity 
of the body, $2\bTa\bPsi$ shares the same uniaxial symmetry of $\bTz$ and takes the form of a dipole interaction.} 
Other choices are of course possible and the specific form of 
$\bTa$ has to be inferred by the interaction of a number of features, namely, 
the properties of the particular material under study, the experimental results 
and microscopic theoretical investigations. 
\comm{Instead of an isotropic $\bTa$, a second natural possibility is to choose $\bTa \propto \bPsi$. However, in both 
cases the resulting \emph{active stress}, $2\bTa\bPsi$, possesses the same uniaxial symmetry about the director $\bn$. 
Therefore, both choices should lead to essentially the same qualitative results.}

Finally, the last term in Eq.\eqref{eq:Tdef_active_weak} is standard and it is a consequence of the 
elastic distortions of the director field.

\section{Spontaneous flow and self-channeling}
\label{sec:SpontaneousFlow}
\begin{figure}
\includegraphics[width=0.45\textwidth]{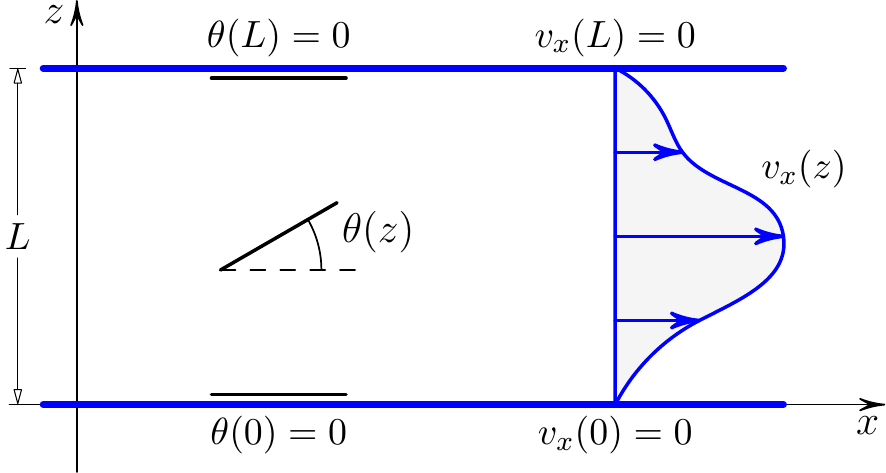}
\caption{Schematic representation of the channel geometry studied in the text.}
\label{fig:channel} 
\end{figure}

A key prediction of continuum models of active liquid crystals is the existence 
of spontaneously generated fluid flows in one dimensional channels 
\cite{05voituriez,08Giomi,09edwards,12Giomi}. It is then natural to test the 
present model against such predictions. To this end, I analyze the hydrodynamic 
equations in the simple geometry of a two-dimensional channel of infinite length 
along the $x$-direction and with height $L$ (see Fig.\ref{fig:channel}).
For simplicity, I also assume that there is only one relaxation time, i.e. $\bbD 
= \tfrac{1}{2}\rho\mu\tau \bbI$, and that $\tau$ is much smaller than the 
characteristic times of the flow ($\bbI$ is the fourth-rank 
identity tensor). Hence, it is possible to use the approximation of small 
effective deformations, as developed in Sec.\ref{sec:weak-flow}, where the 
Cauchy stress tensor is given as in Eq.\eqref{eq:Tdef_active_weak}. I also 
assume incompressibility so that the pressure is now a Lagrange multiplier. I 
consider a two-dimensional active nematic suspension with uniform degree of 
orientation and discuss the spatial dependence of the fields $\theta(z)$ (the 
angle that the nematic units form with the $x$-axis) and $v_x(z)$ (the 
$x$-component of the macroscopic velocity of the suspension). Both fields depend 
only on $z$ because of the translational invariance along $x$. I assume no slip 
boundary conditions at both surfaces and $\theta(0)=\theta(L)=0$. The equations 
of motions in steady conditions are the Stokes equations $\partial 
T_{xz}/\partial z=0$, $\partial T_{zz}/ \partial z=0$ and the director equation 
$\bn\times\bh=0$, namely Eq.\eqref{eq:evoluzione_direttore} with $\bg=0$. The second of these equations determines the 
pressure in the film, and I will not discuss this in more detail. More interesting are the first and the third that 
read \comm{(see Appendix \ref{sec:appendixB} for an explicit derivation)}
\begin{subequations}
\begin{align}
& 4 (a_{0}^3-1) \theta ' \Big[2 \tau  v_{x}' \sin (2 \theta) \left(\left(a_{0}^3-1\right) 
\cos (2 \theta)+a_{0}^3+1\right) \notag \\
& - 2 a_{0} \zeta  \cos (2 \theta) \Big]
-\tau  v_{x}'' \Big[ 4 \left(a_{0}^6-1\right) \cos (2\theta)-5 a_{0}^6 \notag \\
+ & \left(a_{0}^3-1\right)^2 \cos (4 \theta) + 2 a_{0}^3-5 \Big] = 0,
\label{eq:SpFlow_vx}
\end{align}
\begin{equation}
\begin{split}
& \left(a_{0}^3-1\right) \mu  \tau  v_{x}' \left(\left(a_{0}^3+1\right) \cos (2 \theta) - a_{0}^3+1\right) \\
& + 2 a_{0}^2 k \,\theta '' = 0 ,
\end{split}
\label{eq:SpFlow_theta}
\end{equation}
\label{eq:SpFlow}
\end{subequations}
with boundary conditions $\theta(0)=\theta(L)=0$ and $v_x(0)=v_x(L)=0$. The 
coefficient $a_0$ is the shape parameter that identifies the form of the shape 
tensor at the given density $\rho_0$: $a_0 = a(\rho_0)$. The Oseen-Frank 
potential, $\sigma_{\text{Fr}}(\rho,\bn, \nabla\bn)$, is simply taken to be 
$\sigma_{\text{Fr}} = k|\nabla\bn|^2$ (one-constant approximation).

It is straightforward to check that $\theta(z)=0$ and $v_x(z)=0$ is always a 
solution of \eqref{eq:SpFlow}, for any value of the parameters. However, above a 
critical threshold for the thickness $L$, a bifurcation occurs and the trivial 
solution is no longer unique. The critical condition is obtained by performing a 
linear stability analysis. The linearized equations about the trivial solution 
reads
\begin{subequations}
\begin{align}
\tau \, v_x''(z) - \zeta\, a_0 \left(a_0^3-1\right) \theta'(z) & = 0, \\
a_0^2 k \,\theta''(z) + \mu  \tau \left(a_0^3-1\right) v_x'(z) & = 0.
\end{align}
\label{eq:linearized}
\end{subequations}
It is known \cite{05voituriez,09edwards,12Giomi} that both the polar and the 
apolar systems exhibit a pitchfork bifurcation, i.e., a Fre\'{e}dericksz-like 
transition, between a state where the director field is constant and parallel to 
the walls throughout the channel to a nonuniformly oriented state in which the 
system spontaneously flows in the $x$ direction. The transition can be tuned by 
changing either the film thickness or the activity parameter.

However, in our case, the linearized equations \eqref{eq:linearized} admit 
\emph{two} independent modes of instability at the bifurcation and hence 
\eqref{eq:linearized} show a more complex behavior than a simple pitchfork 
bifurcation. Namely, the critical condition reads
\begin{equation}
L \sqrt{\frac{\left(a_0^3-1\right)^2}{a_0} \frac{\mu\zeta}{k}}=2 \pi  n, 
\label{eq:critical}
\end{equation}
where $n$ is an integer. For a given $L$, the lowest critical value of the 
activity, $\zeta_c$, is the one where bifurcation occurs and corresponds to the 
fundamental mode $n=1$. Likewise, for a given activity coefficient $\zeta$, it 
is possible to define a critical length $L_c$ corresponding to the solution of 
\eqref{eq:critical} with $n=1$. It is interesting to observe that $L_c$ diverges 
to infinity, and no bifurcation occurs, either for $\zeta \to 0$ (passive case) 
or $a_0 \to 1$ (isotropic case). When the condition \eqref{eq:critical} is 
satisfied, the non-trivial solutions of Eqs.\eqref{eq:linearized} are
\begin{subequations}
\begin{align}
\theta(z) & = C_1 \sin^2\left(\tfrac{\pi  n z}{L}\right)
+ C_2 \sin \left(\tfrac{2 \pi  n z}{L}\right) , \\
v_x(z) & = \frac{a_0 \left(a_0^3-1\right) \zeta  L}{4 \pi  n \tau }
\Big(4 C_2 \sin ^2\left(\tfrac{\pi  n z}{L}\right) \notag \\
& - C_1 \sin \left(\tfrac{2 \pi  n z}{L}\right)\Big),
\end{align}
\label{eq:linsolutions}
\end{subequations}
with $C_1$ and $C_2$ arbitrary coefficients. Therefore, they span a 2-dimensional linear space for each value of $n>0$.\\

This twofold instability is confirmed by \comm{a numerical analysis. In the 
numerical code}, $L$ is used to set the length-scale, while $\tau$ sets the 
time-scale. The non-dimensional parameters $a_0$ and $\zeta$ are fixed to be 
equal to 2.5 and 0.5, respectively. We then perform the numerical integration of 
the nonlinear equations \eqref{eq:SpFlow}, where the ratio $k/\mu$ is chosen 
such that the critical length $L_c$ can be suitably adjusted. 

\begin{figure}
\includegraphics[width=0.3\textwidth]{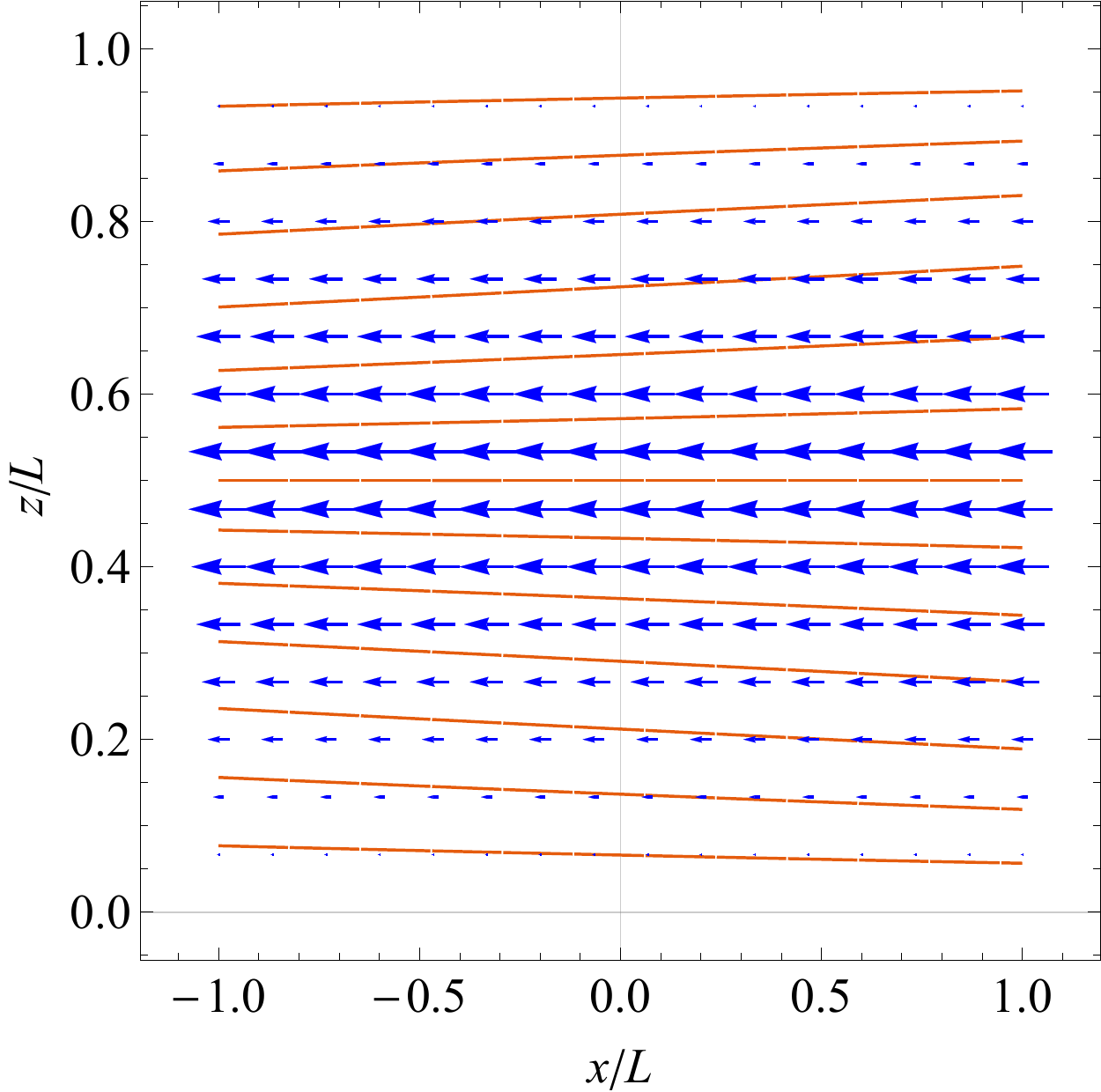} 
\caption{Spontaneous flow. Vector plot of the solutions of 
Eqs.\eqref{eq:SpFlow}, with $a_0=2.5$, $\zeta=0.5$ and $L=1.1\,L_c$, where $L_c$ 
is given as in Eq.\eqref{eq:critical} $(n=1)$. Blue arrows represent the 
velocity field $v_x(z)$, while orange lines depict the director field.}
\label{fig:spflow}
\end{figure}

\begin{figure}
\includegraphics[width=0.3\textwidth]{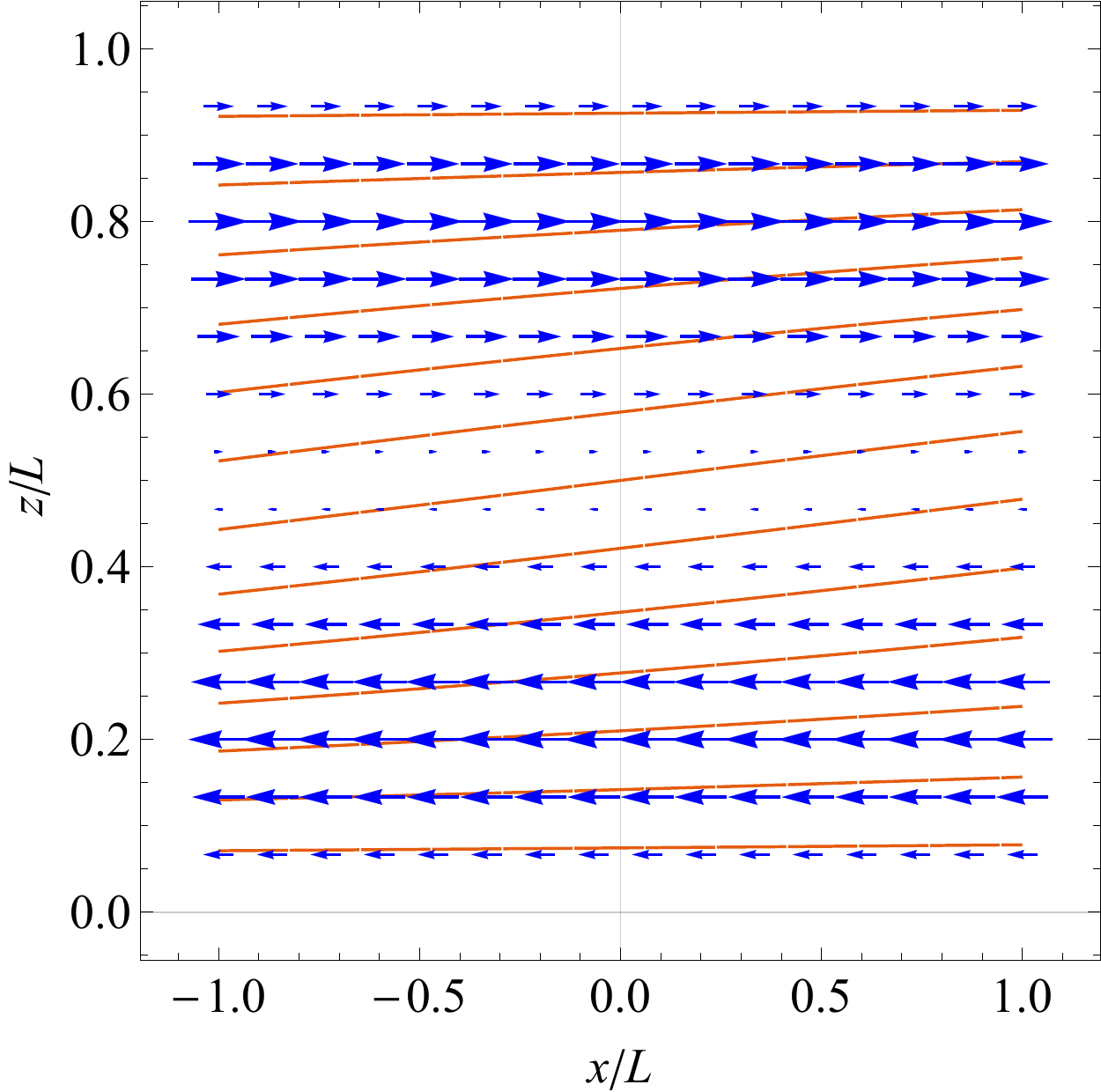} 
\caption{Self-channeling. Vector plot of the solutions of Eqs.\eqref{eq:SpFlow}, 
with $a_0=2.5$, $\zeta=0.5$ and $L=1.4\, L_c$, where $L_c$ is given as in 
Eq.\eqref{eq:critical} $(n=1)$. Blue arrows represent the velocity field 
$v_x(z)$, while orange lines depict the director field.}
\label{fig:selfchannelling}
\end{figure}

Figure \ref{fig:spflow} shows a numerical solution of Eqs. \eqref{eq:SpFlow} 
for $L=1.1\, L_c$. The first mode to be excited above the bifurcation is the 
``spontaneous flow'' mode, corresponding to $C_1=0$ and $C_2 \neq 0$ in 
Eqs.\eqref{eq:linsolutions}. A right or left steady flow of active particles is 
spontaneously generated due to activity. Upon further increasing $L/L_c$, also 
the self-channeling mode, corresponding to $C_1 \neq 0$ and $C_2 = 0$, 
appears. Figure \ref{fig:selfchannelling} shows a numerical solution of 
Eqs.\eqref{eq:SpFlow} with $L=1.4\, L_c$. Here, the active particles 
self-organize into sub-channels and show no net flow of particles.

While a Fre\'{e}dericksz-like transition from a planar motionless state to a 
flowing state has been predicted for increasing thickness, in an active 
\emph{nematic} gel the transition to a self-channeling state seems to be new. A 
similar effect has been described in \cite{12Giomi}, but in the case of a 
two-dimensional square domain with periodic boundary conditions. The authors of 
Ref.\cite{12Giomi} do not observe any banding in the simple one-dimensional 
geometry considered here. By contrast, active \emph{polar} gels seems to show a 
richer behavior where self-channeling appears \cite{05voituriez,09edwards}.\\

\section{Conclusion}
Active nematic gels are non-conventional materials that mimic the behavior of 
living matter. They represent an excellent playground for a deeper understanding 
of the mechanical response of biological tissues such as the cellular 
cytoskeleton, a network of cross-linked filaments subjected to the action of 
molecular motors. 

The simplest theory of active nematic gels takes inspiration from liquid crystal 
theory, which is then typically supplemented by adding an active stress and a 
viscoelastic relaxation time to the passive theory. However, this model does not 
explicitly take into account the fact that molecular motors act at the 
microscopic level by modifying the way in which the material reorganize its 
internal structure. 

To this end, I have put forth a thermodynamically consistent theory of active 
nematic gels that naturally embeds viscoelasticity and introduces activity as a 
remodeling force so that no active stress has to be added to the model. The 
active remodeling force competes with the natural relaxation process of the 
passive systems and drives the system out of equilibrium. 

Finally, in order to explore the early consequences of the theory, I have 
studied the dynamical properties of thin films of active nematic fluids.  Above 
a critical thickness of the film a rich variety of complex behaviors is 
observed. Namely, at the same critical length, I find both a spontaneous flow 
of active particles and a self-channelling effect, where the particles organize 
themselves into sub-channels and flow in opposite directions. 

\begin{acknowledgments} I wish to thank Davide Ambrosi and Antonio Di Carlo for 
their encouragement and for stimulating discussions.

\end{acknowledgments}

\input{ActiveNematics_Rev.bbl}

\appendix
\section{Derivation of the Cauchy stress tensor and the molecular field}
\label{sec:appendixA}
We need two simple lemmas, which I state without proofs.
\begin{lemma} Let $f$ be a function depending only on $\bFe$ (or $\bBe$). Then,
\begin{enumerate}
\item[(1)] $\displaystyle\D{f}{\bF}\bF^{T} = \D{f}{\bFe}\bFe^{T} = 2\D{f}{\bBe}\bBe$
\item[(2)] $\displaystyle\bF^{-T}\D{f}{\bH}\bF^{-1} = \D{f}{\bBe}$
\end{enumerate}
\label{teo:lemma1}
\end{lemma}

\begin{lemma} Let $f=f(\rho)$ be a function of $\rho$ only. Then, $\displaystyle\D{f}{\bF}\bF^{T} = -\rho\D{f(\rho)}{\rho} \bI$.
\label{teo:lemma2}
\end{lemma}

We can now calculate the material derivative of $\FF$
\begin{align}
\dot{\FF} 
& = \int_{\Pt} \Big(\rho\D{\si}{\bF}\cdot \Ldot{\bF} + \rho\D{\si}{\bH}\cdot \Ldot{\bH}
+ \rho \D{\si}{\bn}\cdot \Ldot{\bn} \notag \\
&\qquad \qquad  + \rho \D{\si}{\nabla\bn}\cdot (\nabla\bn)\dot{\phantom{i}} \Big) dv.
\end{align}
which is then simplified with the use of Lemma \ref{teo:lemma1} and the identities
\begin{align}
\Ldot{\bF} & = (\nabla \bv)\bF, \\
\frac{D}{Dt} (\nabla\bn) & = \nabla \Ldot{\bn} - (\nabla \bn)(\nabla \bv), \\
\bF\Ldot{\bH}\bF^{T} & = (\bBe)\Ldot{\phantom{I}} - (\nabla\bv) \,\bBe - \bBe \,(\nabla\bv)^T =: \bBe^{\uctd},
\end{align}
so that the rate of change of the free energy reads 
\begin{align}
\dot{\FF} & = \int_{\Pt} \left(\rho\D{\si}{\bF}\bF^{T} - \rho(\nabla \bn)^T\D{\si}{\nabla\bn} \right) \cdot \nabla \bv \,\,dv \notag \\
& + \int_{\Pt} \left(\rho\D{\si}{\bn} +\rho\D{\si}{\nabla\bn} \cdot \nabla\Ldot{\bn}\right) \,\, dv \notag \\
& + \int_{\Pt}\rho\D{\si}{\bBe}\cdot \bBe^{\uctd}\,\, dv.
\end{align}
Further simplifications are obtained by employing the divergence theorem  in the first two integrals
\begin{align}
\int_{\Pt} \bX \cdot \nabla \bv \,\,dv & = \int_{\partial\Pt} \bX\bnu \cdot \bv \,\,da \notag \\
& - \int_{\Pt} \divr(\bX) \cdot \bv \,\,dv, 
\end{align}

\begin{align}
\int_{\Pt} \bX \cdot \nabla \Ldot{\bn} \,\,dv & = \int_{\partial\Pt} \bX\bnu \cdot \Ldot{\bn} \,\,da \notag \\
& - \int_{\Pt} \divr(\bX) \cdot \Ldot{\bn} \,\,dv,
\end{align}
where $\bX$ is a generic second-rank tensor. In so doing, we recognize the 
generalized forces paired to the fields $\bv$ and $\Ldot{\bn}$ and obtain 
\begin{align}
\dot{\FF} & = \int_{\partial\Pt} \left(\rho\D{\si}{\bF}\bF^{T} - \rho(\nabla \bn)^T\D{\si}{\nabla\bn} \right)\bnu \cdot \bv \,\,da \notag \\
& - \int_{\Pt} \divr\left(\rho\D{\si}{\bF}\bF^{T} - \rho(\nabla \bn)^T\D{\si}{\nabla\bn} \right) \cdot \bv \,\,dv \notag \\
& + \int_{\partial\Pt} \left(\rho\D{\si}{\nabla\bn}\right) \bnu \cdot \Ldot{\bn}\,\, da \notag \\
& + \int_{\Pt} \left[\rho\D{\si}{\bn} - \divr\left(\rho\D{\si}{\nabla\bn} \right)\right] \cdot\Ldot{\bn} \,\, dv \notag \\
& + \int_{\Pt}\rho\D{\si}{\bBe}\cdot \bBe^{\uctd}\,\, dv.
\label{eq:dotF}
\end{align}

It is then natural to define the Cauchy stress tensor (conjugate to $\bv$) and 
the molecular field (conjugate to $\Ldot{\bn}$)
\begin{align}
\bT &:= \rho\D{\si}{\bF}\bF^{T} - \rho(\nabla \bn)^T\D{\si}{\nabla\bn}, 
\label{eq:T_F} \\
\bh &:= \rho\D{\si}{\bn} - \divr\left(\rho\D{\si}{\nabla\bn} \right).
\end{align}
A further application of Lemmas \ref{teo:lemma1} and \ref{teo:lemma2} allow us 
to rewrite the stress tensor \eqref{eq:T_F} as in Eq.\eqref{eq:Tdef}.

\comm{
\section{Derivation of Eq.\eqref{eq:SpFlow}}
\label{sec:appendixB}
\newcommand{\bnp}{\bn^{\perp}}

Let us denote with $\{\be_x,\be_y,\be_z\}$ the Cartesian unit vectors along the coordinate axes. We posit a stationary 
velocity field of the form $\bv= v_x(z) \be_x$. Hence, the gradient of velocity is 
$\nabla\bv = v_x'(z) \be_x \tp \be_z$, and the material time-derivative of $\bv$ vanishes. The director field is 
described by the angle $\theta(z)$ such that $\bn = \cos\theta(z)\,\be_x + \sin\theta(z)\,\be_z$. It is also useful to 
introduce the orthogonal unit vector $\bnp = -\sin\theta(z)\,\be_x + \cos\theta(z)\,\be_z$.
With the simplifying assumptions
\begin{equation}
\bTa = -\frac{1}{2}\rho\mu \zeta \bI, \qquad \bbD = \tfrac{1}{2}\rho\mu\tau \bbI, 
\end{equation}
the stress tensor, as given in Eq.\eqref{eq:Tdef_active_weak}, reads
\begin{equation}
\bT = -p \bI - \rho\mu \left(\tau \bPsi^{\uctd}\bPsi + \zeta \bPsi \right) - \rho k (\nabla\bn)^T (\nabla\bn),
\label{eq:T_appendix}
\end{equation}
where $\nabla\bn = \theta'(z) \,\bnp \tp \be_z$, so that 
\begin{equation}
(\nabla\bn)^T (\nabla\bn) = \theta'(z)^2 \,\be_z \tp \be_z. 
\end{equation}
To calculate the codeformational derivative of the shape tensor, we observe that $\D{\bPsi}{t} = 0$ and $ 
(\nabla\bPsi)\bv = 0$ so that we obtain
\begin{align}
\bPsi^{\uctd} & = \D{\bPsi}{t} + (\nabla\bPsi)\bv - (\nabla\bv)\bPsi - \bPsi (\nabla\bv^T) \notag\\
& = -\left(a_0^2 - \frac{1}{a_0}\right) v_x'(z) \sin\theta(z) \left(\be_x \tp \bn + \bn \tp \be_x \right) \notag \\
& - \frac{1}{a_0} v_x'(z) \left(\be_x \tp \be_z + \be_z \tp \be_x \right),
\end{align}
where we have assumed incompressibility with constant density $\rho_0$ and $a(\rho_0) = a_0$, so that the pressure $p$ 
is a Lagrange multiplier. This derivative is then inserted into Eq.\eqref{eq:T_appendix} to obtain, with lengthy but 
straightforward calculations, the Cauchy stress tensor. 

Since all the terms depend only on the variable $z$, the Stokes equations reduce to 
\begin{equation}
\D{T_{xz}}{z} = 0, \qquad \D{T_{zz}}{z} = 0.
\end{equation}
The latter yields and equation for the pressure, while the former is
\begin{align}
\frac{d}{dz}&\Big[-\frac{\rho\mu}{8 a_0^2}  \Big(4 a_0 \left(a_0^3-1\right) \zeta  \sin (2 \theta (z)) \notag \\
& + \tau  v_x'(z) \big(4 \left(a_0^6-1\right) \cos (2 \theta (z)) - 5 a_0^6 \notag \\
& + \left(a_0^3-1\right)^2 \cos (4 \theta (z)) + 2 a_0^3-5\big)\Big) \Big] = 0,
\end{align}
and corresponds to Eq.\eqref{eq:SpFlow_vx}.

The evolution equation of the director furnishes Eq.\eqref{eq:SpFlow_theta}. In order to show this, let us calculate 
the molecular field $\bh$, as given in Eq.\eqref{eq:hdef}. To this end, we observe that
\begin{align}
\D{\sigma}{\bn} & = 2(a_0^2-a_0^{-1})\D{\sigma}{\bPsi}\bn, \\
\D{}{\bPsi} \tr(\bPsi^{-1}\bBe) & = -\bPsi^{-1}\bBe\bPsi^{-1}, \\
\D{\sigma}{\bn} & = -\mu(1-a_0^{-3})\bPsi^{-1}\bBe\bn, \\
\divr \D{\sigma}{\nabla\bn} & = k \divr\nabla\bn = k(\theta''\, \bnp - \theta'^2\bn).
\end{align}
In agreement with Eq.\eqref{eq:T_appendix}, the term $\bPsi^{-1}\bBe$ is \begin{equation}
\bPsi^{-1}\bBe = \bI - \tau \bPsi^{\uctd}\bPsi - \zeta \bPsi,
\end{equation}
and the director equation, $\bn \times \bh = 0$, then reads
\begin{equation}
\bn \times \Big( \mu \tau (a_0^2-a_0^{-1})\, \bPsi^{\uctd}\bn - k \theta''\, \bnp \Big) = 0.
\end{equation}
After a further simplification, this coincides with Eq.\eqref{eq:SpFlow_theta}.
}

\end{document}

%% file: ActiveNematics_Rev.bbl
%